\documentclass[11pt,twoside]{article}
\usepackage{asp2004}
\usepackage{psfig}
\usepackage{epsf}
\usepackage{graphics}
\usepackage{lscape}

\begin{document}
\title{XMM-Newton Observations of Recent Galactic Novae}
\author{Margarita Hernanz and Gl\`oria Sala}
\affil{Institut d'Estudis Espacials de Catalunya (CSIC). Campus UAB, 
Facultat de Ci\`encies, Torre C5-parell, 2a planta.
E-08193 Bellaterra (Barcelona), Spain}

\begin{abstract}
We report on the observations of five recent galactic classical novae,
performed with XMM-Newton. Some of them show emission 
in the whole range from 0.2 to 8.0 keV, up to 3 years after the
explosion. Nova Oph 1998 shows clear 
evidence of resumed accretion less than 3 years after its
explosion, indicating a very fast recovery of the cataclysmic variable after 
the nova outburst. In another case, Nova Sgr 1998, it is difficult
to disentangle if the emission is produced in the ejecta and/or 
in the accretion stream.
\end{abstract}

\section*{Introduction}
Five classical novae have been observed with XMM-Newton between two
and four years after the outburst. Two of them are bright X-ray sources, 
with their X-ray emission spanning
the whole energy range of the EPIC cameras, 0.2-8 keV. In both
cases, the spectrum indicates that the main origin of the X-rays is
not photospheric, and that hydrogen burning has ceased on the white
dwarf envelope, thus indicating turn-off times shorter than $\sim 3$ 
years.

\section*{Nova Sgr 1998 (V4633 Sgr)}

The X-ray spectrum of Nova Sgr 1998 (V4633 Sgr) is dominated by thermal
plasma emission, which can be due to the shock-heated expanding shell,
or to the reestablishment of accretion in the cataclysmic variable.
The best-fit model is a three-temperature thermal plasma, with temperatures
between 0.1 keV and 40 keV. Thermal plasma models with
different compositions (solar, and several CO and ONe enhanced from
realistic nova models) have been tried, and the best-fit is obtained
either with a CO enhanced model, or with the solar abundances. The
first case would correspond to emission from the expanding nova shell.
The unabsorbed luminosity of the thermal plasma component is indeed
compatible with the X-ray emission from other observed nova shells,
as well as the electronic densities derived from the emission measure.
The possibility of accretion being reestablished, which would be simulated
by the solar thermal plasma model, can not be ruled out. In this case,
the luminosity of the plasma would be compatible with accretion in
a intermediate polar cataclysmic variable. Optical
observations of V4366 Sgr also support the magnetic character of this
nova, indicating that it is a nearly synchronous system \citep{lip01}.

\section*{Nova Oph 1998 (V2487 Oph)}
XMM-Newton observations of Nova Oph 1998 (V2487 Oph) surprised us
showing that accretion had been reestablished on the cataclysmic variable
less than 1000 days after outburst \citep{HS02}. The spectrum
in this case is also dominated by thermal plasma emission, but an
additional excess around 6.4 keV indicates the presence of
iron K$ _\alpha$ fluorescent lines. Fluorescent lines 
arise from reflection of hard X-rays on a cold plasma, thus  
implying the reestablishment of
accretion by the time of observations. The spectrum shows also a soft
component associated to a fraction of the white dwarf surface heated
by hard X-rays from the accretion shock region, which is another common
feature in accreting cataclysmic variables, specially on magnetic
systems. The high temperature and large luminosity of the thermal
plasma emission also indicate that V2487 Oph probably occurred on
a magnetic white dwarf.

V2487 Oph is twice exceptional, since the cataclysmic variable hosting
the nova was also detected eight years before the outburst during
the ROSAT All Sky Survey. This detection makes V2487 Oph the first
classical nova detected in X-rays before and after the outburst, and
supports the cataclysmic variable scenario for classical novae.

\begin{figure}
\plottwo{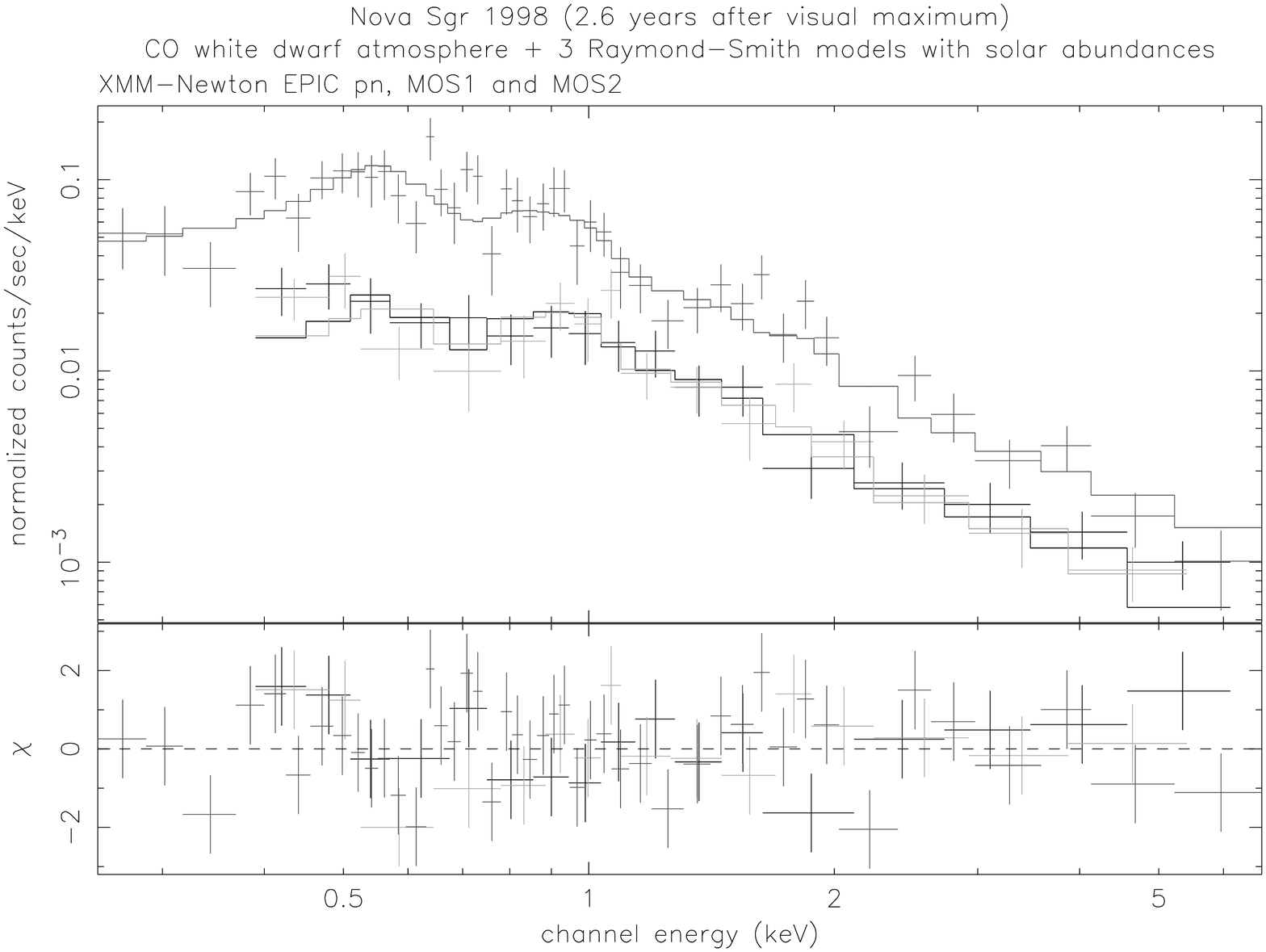}{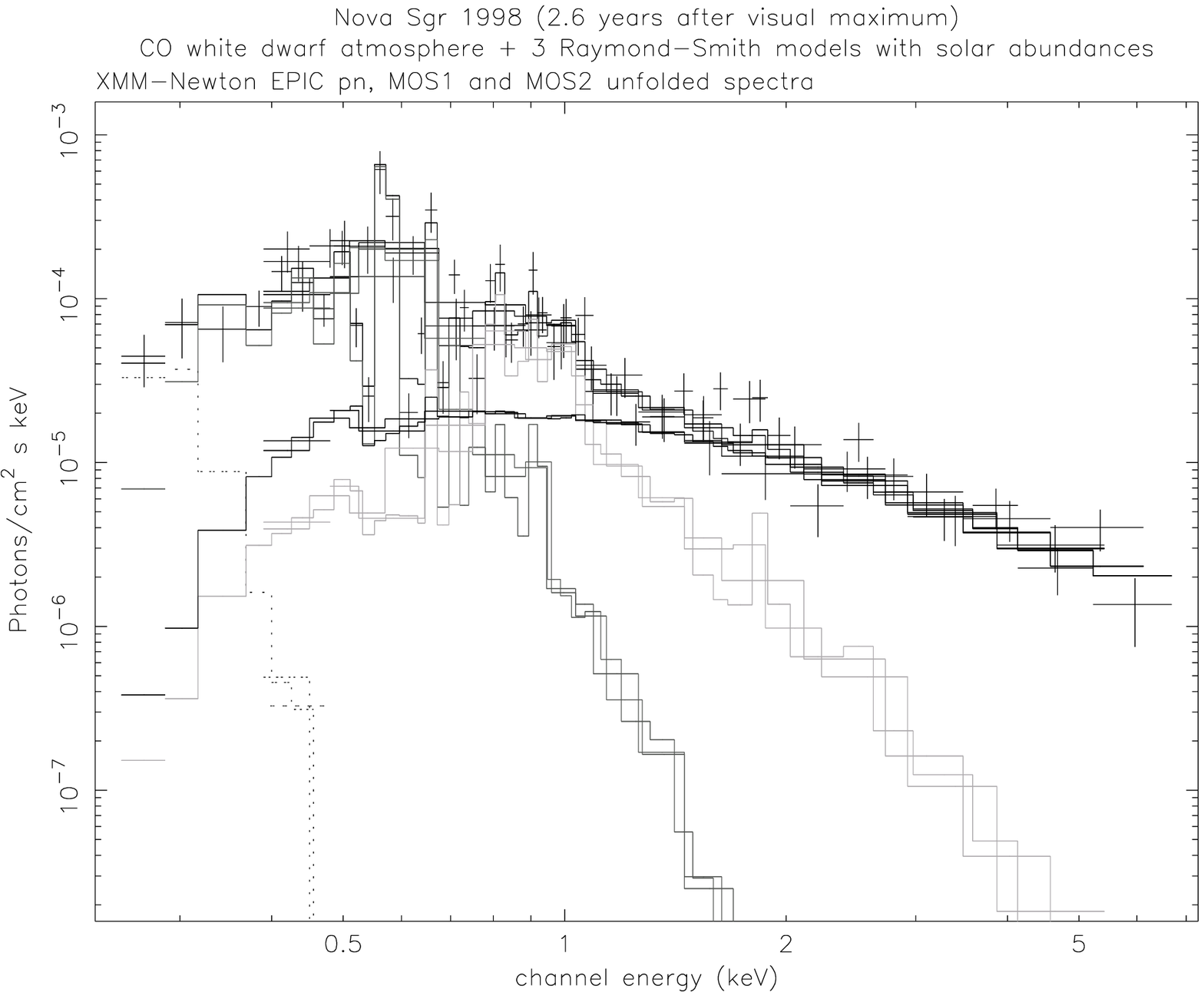}
\caption{\textbf{Left panel:} Observed X-ray spectrum for V4633 Sgr 
and residuals. Data from the three EPIC cameras are
indicated: pn (black thin line, with higher count-rate), MOS1 (black thick line) and MOS2
(gray). The best-fit model is a CO white dwarf atmosphere plus a 3-temperature 
Raymond-Smith with solar abundances.
\textbf{Lower panel:} Unfolded spectrum, showing the contribution of each component. From left to right,
CO white dwarf atmosphere and Raymond-Smith models
with kT$_{1}$=0.11 keV, kT$_{2}$=0.82 keV and  kT$_{3}$=37 keV.}

\label{figures}
\end{figure}

\end{document}